\documentstyle[12pt,aps,epsf,tighten,epsfig,amsmath,multicol,prbbib]
 {revtex}

\normalsize

\def\bG{{\bf G}}
\def\bi{{\bf i}}
\def\bj{{\bf j}}
\def\bk{{\bf k}}

\def\bq{{\bf q}}
\def\bG{{\bf G}}

\def\b0{{\bf 0}}

\def\tF{\tilde F}
\def\tG{\tilde G}
\def\tH{\tilde H}

\def\tS{\tilde S}

\def\tPi{\tilde\Pi}
\def\tSg{\tilde\Sigma}
\def\txi{\tilde\xi}
\def\btG{{\bf\tG}}
\def\btSg{{\boldsymbol\tSg}}

\def\bra{\langle}
\def\ket{\rangle}
\def\up{\uparrow}
\def\down{\downarrow}

\def\eps{\epsilon}

\def\Lam{\Lambda}
\def\om{\omega}
\def\sg{\sigma}

\begin{document}

\title{Renormalized perturbation theory for Fermi systems: \\
 Fermi surface deformation and superconductivity in the two-dimensional
 Hubbard model}

\author{Arne Neumayr \\
{\em Institut f\"ur Theoretische Physik C, Technische Hochschule Aachen} \\
{\em D-52056 Aachen, Germany}}

\author{Walter Metzner \\
{\em Max-Planck-Institut f\"ur Festk\"orperforschung, D-70569 Stuttgart, 
 Germany}}

\date{\small\today}
\maketitle


\begin{abstract}
Divergencies appearing in perturbation expansions of interacting 
many-body systems can often be removed by expanding around a 
suitably chosen renormalized (instead of the non-interacting) 
Hamiltonian.
We describe such a renormalized perturbation expansion for 
interacting Fermi systems, which treats Fermi surface shifts 
and superconductivity with an arbitrary gap function via 
additive counterterms.
The expansion is formulated explicitly for the Hubbard model 
to second order in the interaction. 
Numerical solutions of the self-consistency condition determining
the Fermi surface and the gap function are calculated for the
two-dimensional case. 
For the repulsive Hubbard model close to half-filling we find
a superconducting state with d-wave symmetry, as expected.
For Fermi levels close to the van Hove singularity a Pomeranchuk 
instability leads to Fermi surfaces with broken square lattice 
symmetry, whose topology can be closed or open.
For the attractive Hubbard model the second order calculation
yields s-wave superconductivity with a weakly momentum dependent
gap, whose size is reduced compared to the mean-field result.

\noindent
\mbox{PACS: 71.10.Fd, 71.10.-w, 74.20.Mn} \\
\end{abstract}

\section{Introduction} 

Unrenormalized perturbation expansions of interacting electron
systems around the non-interacting part of the Hamiltonian are
generally plagued by infrared divergencies. 
Some of the divergencies are simply due to to shifts of the Fermi 
surface, while others signal instabilities of the normal Fermi 
liquid towards qualitatively different states, such as 
superconducting or other ordered phases.
This problem is often treated by self-consistent resummations 
of Feynman diagrams, where a finite or infinite subset of
skeleton diagrams, with the interacting propagator $G$ on 
internal lines, is summed.\cite{BK} 
Symmetry breaking can be built into the structure of $G$ as an 
ansatz, and the size of the corresponding order parameter is 
determined self-consistently.
This standard approach has been very useful in many cases.
However, resummation schemes beyond
first order (Hartree-Fock) require extensive numerics, since
the full self-energy has to be determined self-consistently,
and delicate low-energy structures cannot always be resolved.
A more serious problem is the fact that self-energy and vertex 
corrections are not treated on equal footing in most feasible 
resummation schemes. This often leads to unphysical results.

In this work we will describe and apply an alternative procedure, 
which has been formulated already long ago by Nozi\`eres,\cite{Noz}
and more recently been discussed in the mathematical literature 
as a way of carrying out well-defined perturbation expansions for
weakly interacting Fermi systems.\cite{FMRT,FKST}
The basic idea is to choose an improved starting point for the 
perturbation expansion, by adding a suitable counterterm to the
non-interacting part of the Hamiltonian, and subtracting it from
the interaction part. The counterterm is quadratic in the Fermi
operators and has to be determined from a self-consistency
condition.
In Sec.\ II we will describe how Fermi surface deformations and 
superconductivity can be treated by this method.
Explicit expressions up to second order in the interaction are
derived for the case of the Hubbard model in Sec.\ III.
Results obtained from a numerical solution of the self-consistency 
equations in two dimensions will follow in Sec.\ IV. 
For the repulsive Hubbard model we have obtained superconducting 
solutions with d-wave symmetry in agreement with widespread
expectations,\cite{Sca} and with recent renormalization group 
calculations which conclusively established d-wave 
superconductivity at weak coupling.\cite{RG}
In addition, for Fermi levels close to the van Hove singularity, 
deformations which break the square lattice symmetry occur.
This confirms the recently proposed possibility of 
symmetry-breaking Fermi surface deformations 
(''Pomeranchuk instabilities'').\cite{HM,YK,GKW}

\section{Renormalized perturbation expansion} 

We consider a system of interacting spin-$\frac{1}{2}$ fermions
with a Hamiltonian $H = H_0 + H_I$, where the non-interacting
part
\begin{equation}
 H_0 = \sum_{\bk,\sg} \xi_{\bk} \, n_{\bk\sg}
\end{equation}
with $\xi_{\bk} = \eps_{\bk} - \mu$ contains the kinetic energy
and the chemical potential, while $H_I$ is a fermion-fermion 
interaction term. 
We are particularly interested in lattice systems,
for which the dispersion relation $\eps_{\bk}$ is not isotropic.
We consider only ground state properties, that is the temperature
is zero throughout the whole article.

The bare propagator in a standard many-body perturbation expansion
\cite{NO} around $H_0$ is given by
\begin{equation}
 G_0(k) = \frac{1}{i\om - \xi_{\bk}} \; ,
\end{equation}
where $\om$ is the Matsubara frequency and $k = (\om,\bk)$.
This progagator diverges for $\om \to 0$ and $\bk \to \bk_F$,
for any Fermi momentum $\bk_F$, since $\xi_{\bk_F} = 0$.
As a consequence, many Feynman diagrams diverge. 
A well-known singularity is the (usually) logarithmic divergency
of the 1-loop particle-particle contribution to the two-particle 
vertex in the Cooper channel, which leads to a $(\log)^n$
divergency of the n-loop particle-particle ladder diagram.
This signals a possible Cooper instability towards 
superconductivity. Much stronger divergencies occur in
diagrams with multiple self-energy insertions on the same
internal propagator line, leading to non-integrable powers of 
$G_0(k)$.\cite{FMRT,FKST}
These singularities are due to Fermi surface shifts generated 
by the interaction term in the Hamiltonian.

The divergency problems and the superconducting instability
can be treated by splitting the Hamiltonian in a different way,
namely as\cite{Noz}
\begin{equation}
 H = \tH_0 + \tH_I \; ,
\end{equation}
where $\tH_0 = H_0 + \delta H_0$ and $\tH_I = H_I - \delta H_0$,
and expanding around $\tH_0$.
The {\em counterterm}\/ $\delta H_0$ must be quadratic in the 
creation and annihilation operators to allow for a straightforward 
perturbation expansion based on Wick's theorem.
It is possible to chose $\delta H_0$ such that $\tH_I$ does not 
shift the Fermi surface corresponding to $\tH_0$ any more, 
and divergencies due to self-energy insertions are removed.
In the superconducting state spontaneous symmetry breaking can
be included already in $\delta H_0$, with an order parameter
$\Delta_{\bk}$ whose value on the Fermi surface is not shifted
by $\tH_I$.
We will now describe this procedure in more detail.

\subsection{Normal state}

A counterterm 
$\delta H_0 = \sum_{\bk,\sg} \delta\xi_{\bk} \, n_{\bk\sg}$
leads to a renormalized dispersion relation
$\txi_{\bk} = \xi_{\bk} + \delta\xi_{\bk}$ in the unperturbed
part of the Hamiltonian, 
\begin{equation}
 \tH_0 = \sum_{\bk,\sg} \txi_{\bk} \, n_{\bk\sg} \; ,
\end{equation}
and correspondingly to a new bare propagator
\begin{equation}
 \tG_0(k) = \frac{1}{i\om - \txi_{\bk}} \; .
\end{equation}
The Fermi surface $\tilde{\cal F}$ associated with $\tH_0$ is given by 
the momenta $\tilde\bk_F$ satisfying the equation $\txi_{\bk} = 0$.
The Fermi surface of the interacting system is given by the
solutions of the equation $G^{-1}(0,\bk) = 0$.
This surface coincides with the unperturbed one, corresponding to 
$\tH_0$, if the {\em renormalized}\/ self-energy 
$\tSg = \tG_0^{-1} - G^{-1}$
vanishes on $\tilde {\cal F}$, that is if
\begin{equation}
 \tSg(0,\bk) = 0 \quad \mbox{for} \quad \bk \in {\tilde {\cal F}}
 \; .
\end{equation}
This imposes a self-consistency condition on the counterterms
which can be solved iteratively.
For isotropic systems the shift of $\xi_{\bk}$ can be chosen 
as a momentum independent constant, which may be interpreted 
as a shift of the chemical potential. 
For anisotropic systems, however, one generally has to adjust 
the whole shape of the Fermi surface.
That this procedure really works at each order of the perturbation
expansion has been shown rigorously for a large class of 
systems.\cite{FST}

The shift function $\delta\xi_{\bk}$ is uniquely determined by
the self-consistency condition only on the (interacting)
Fermi surface $\tilde {\cal F}$. 
For momenta away from the Fermi surface, $\delta\xi_{\bk}$ can 
be chosen to be any sufficiently smooth function of $\bk$ which 
does not lead to artificial additional zeros of $\txi_{\bk}$.

The perturbation expansion of the renormalized self-energy $\tSg$ 
involves two types of vertices:
the usual two-particle vertex given by the interaction $H_I$ and
a one-particle vertex due to the counterterm $-\delta H_0$ in $\tH_I$.
In Fig.\ 1 we show the Feynman diagrams contributing to $\tSg$
up to second order in the interaction.

The above-mentioned divergencies of Feynman diagrams with
self-energy insertions on internal propagator lines are removed in
the renormalized expansion around $\tH_0$, since in products
$\tG_0 \tSg \tG_0 \dots \tSg \tG_0$ only one simple pole at
$k=(0,\tilde\bk_F)$ survives, all other poles being
cancelled by the corresponding zeros of the self-energy $\tSg$.

\subsection{Superconducting state}

To treat superconducting states we also add counterterms 
containing Cooper pair creation and annihilation operators, 
in addition to a shift of $\xi_{\bk}$. 
We consider only spin singlet pairing, but triplet pairing could
be dealt with analogously.
We thus expand around a BCS mean-field Hamiltonian
\begin{equation}
 \tH_0 = \sum_{\bk,\sg} \txi_{\bk} \, n_{\bk\sg} +
 \sum_{\bk} \left[ 
 \Delta_{\bk} \, a^{\dag}_{-\bk\down} a^{\dag}_{\bk\up} +
 \Delta^*_{\bk} \, a_{\bk\up} a_{-\bk\down} \right] \; ,
\end{equation}
where $\Delta_{\bk}$ is the gap-function, which has to be 
determined self-consistently.
In terms of Nambu operators
\begin{equation}
 \Psi_{\bk} = \left( \begin{array}{l}
  a_{\bk\up} \\ a^{\dag}_{-\bk\down} \end{array} \right) 
 \quad \mbox{and} \quad
 \Psi^{\dag}_{\bk} = \left(
  a^{\dag}_{\bk\up} \, , \, a_{-\bk\down} \right)
\end{equation}
one can rewrite $\tH_0$ in more compact form as
\begin{equation}
 \tH_0 = \sum_{\bk} \txi_{\bk} \, 
 \Psi^{\dag}_{\bk} \, \sg_3 \, \Psi_{\bk} -
 \sum_{\bk} \Psi^{\dag}_{\bk} \,
 ( \Delta'_{\bk} \, \sg_1 - \Delta''_{\bk} \, \sg_2 ) \, \Psi_{\bk}
 \; ,
\end{equation}
where $\sg_1,\sg_2,\sg_3$ are the Pauli matrices, and $\Delta'_{\bk}$
($\Delta''_{\bk}$) is the real (imaginary) part of $\Delta_{\bk}$. 
The two expressions (7) and (9) for $\tH_0$ differ by the constant 
(c-number) $\sum_{\bk} \txi_{\bk}$, which must be taken into account 
only when absolute energies are computed.
The bare Nambu matrix propagator 
${\bf\tG}_0 = - \bra \Psi \Psi^{\dag} \ket_{\tilde 0}$ 
following from $\tH_0$ is given by
\begin{equation}
 \btG_0^{-1}(k) = \left( \begin{array}{cc}
 i\om - \txi_{\bk} & \Delta_{\bk} \\
 \Delta^*_{\bk} & i\om + \txi_{\bk} \end{array} \right) \; .
\end{equation}
Extending the self-consistency condition for the normal state,
we now require that the matrix self-energy 
$\btSg = \btG_0^{-1} - \bG^{-1}$
vanishes on the Fermi surface (defined by $\txi_{\bk} = 0$),
that is
\begin{equation}
 \btSg(0,\bk) = 0 \quad \mbox{for} \quad \bk \in {\tilde {\cal F}}
 \; .
\end{equation}
Thus, for $\om=0$ and $\bk$ on the Fermi surface, neither the 
diagonal nor the off-diagonal elements of $\btG_0^{-1}(k)$ are 
shifted by the interaction term $\tH_I$. The Feynman diagrams
in Fig.\ 1 apply also to the superconducting case, if lines 
are interpreted as Nambu matrix propagators.

The above renormalized perturbation theory is reminiscent of 
the perturbation theory for symmetry broken phases formulated by 
Georges and Yedidia,\cite{GY} where an order-parameter-dependent
free energy function is constructed by adding Onsager reaction 
terms to the mean field contributions, and the actual order 
parameter is determined by minimizing this free energy.

\section{Application to the Hubbard model}

In this section we derive explicit expressions for the self-energy 
and the counterterms for the ground state $(T=0)$ of the Hubbard 
model, up to second order in the renormalized perturbation 
expansion.

The one-band Hubbard model \cite{Mon}
\begin{equation}
 H = \sum_{\bi,\bj} \sum_{\sg} t_{\bi\bj} \, 
 c^{\dag}_{\bi\sg} c_{\bj\sg} + 
 U \sum_{\bj} n_{\bj\up} n_{\bj\down} -
 \mu N
\end{equation}
describes lattice electrons with a hopping amplitude $t_{\bi\bj}$
and a local interaction $U$. 
Here $c^{\dag}_{\bi\sg}$ and $c_{\bi\sg}$ are creation and 
annihilation operators for electrons with spin projection $\sg$ on a 
lattice site $\bi$, and $n_{\bj\sg} = c^{\dag}_{\bi\sg} c_{\bi\sg}$.
Note that we have included the term $\mu N$ with the total particle
number operator $N$ in our definition of $H$.
The non-interacting part of $H$ can be written in momentum space
as $H_0 = \sum_{\bk} \xi_{\bk} \, n_{\bk\sg}$ where 
$\xi_{\bk} = \eps_{\bk} - \mu$ and $\eps_{\bk}$ is the Fourier
transform of $t_{\bi\bj}$.

Our numerical results will be given for the Hubbard model on a 
square lattice with a hopping amplitude $-t$ between nearest 
neighbors and a much smaller amplitude $-t'$ between
next-nearest neighbors. 
The corresponding dispersion relation is
\begin{equation}
 \eps_{\bk} = -2t(\cos k_x + \cos k_y) - 4t' \cos k_x \cos k_y
 \; .
\end{equation}
We now derive expressions for the self-energy and the
resulting self-consistency equations up to second order in $U$.

\subsection{Normal state}

\subsubsection{First order}

To first order in $U$ the self-energy is obtained as
\begin{equation}
 \tSg^{(1)}(k) = U \int_{k'} \tG_0(k') \, e^{i\om' 0^+}
 \, - \, \delta\xi_{\bk} \; ,
\end{equation}
where $\int_k$ is a short-hand notation for the frequency and
momentum integral, including the usual factors of $(2\pi)^{-1}$
for each integration variable.
The first term results from diagram (1a) in Fig.\ 1, the second 
from diagram (1b). Note that the tadpole diagram (1a) yields a 
k-independent contribution, since the Hubbard interaction is local.
The self-consistency condition (6) for $\tSg^{(1)}$ yields,
after carrying out the $\om'$-integration,
\begin{equation}
 \delta\xi_{\bk} = 
 U \int \frac{d^dk'}{(2\pi)^d} \, 
 \Theta(\mu - \eps_{\bk'} - \delta\xi_{\bk'}) \; ,
\end{equation}
to be satisfied (at least) for $\bk \in \tilde{\cal F}$. Since the 
right hand side of this condition is a constant, it is natural to 
define $\delta\xi_{\bk}$ by this constant for all $\bk$. 
Using Luttinger's theorem one can identify the above momentum
integral with the particle density per spin, such that
$\delta\xi_{\bk} = Un/2$, where $n$ is the total density. 
The self-consistency condition thus yields the $n(\mu)$ relation 
of the interacting system. 
Since the counterterm can be chosen k-independently at first order, 
it may be interpreted as a shift of the chemical potential.

\subsubsection{Second order}

The diagrams (2b) and (2c) from Fig.\ 1 obviously cancel each
other to the extent that the first order diagrams (1a) and (1b) 
cancel. 
Writing $\delta\xi_{\bk} = \delta\xi^{(1)} + \delta\xi^{(2)}_{\bk}$
with $\delta\xi^{(1)}$ given by the constant on the right hand side 
of Eq.\ (15), such that $\delta\xi^{(2)}_{\bk}$ is of order $U^2$
for all $\bk$, the sum of contributions from (2b) and (2c) is of 
order $U^3$ and can thus be ignored at second order. 
Hence, only diagram (2a) contributes to the second order self-energy. 
Using the Feynman rules \cite{NO} one obtains
\begin{equation}
 \tSg^{(2)}(k) = U^2 \int_q \tPi_0(q) \, \tG_0(k-q) \; ,
\end{equation}
where $\tPi_0(q) = - \int_{k'} \tG_0(k') \, \tG_0(k'+q)$.
Adding first and second order terms, one arrives at the second
order self-consistency condition
\begin{equation}
 \delta\xi_{\bk} = 
 U \int \frac{d^dk'}{(2\pi)^d} \, 
 \Theta(-\txi_{\bk'}) \, + \, \tSg^{(2)}(0,\bk) \; .
\end{equation}
The counterterm $\delta\xi_{\bk}$ has to be chosen such that
the above equation is satisfied for all $\bk \in \tilde{\cal F}$,
that is for all $\bk$ satisfying $\txi_{\bk} = 0$.
Since $\tSg^{(2)}(0,\bk)$ is momentum dependent, $\delta\xi_{\bk}$
cannot be chosen constant any more. 
As a consequence, the Fermi surface of the interacting system 
will be deformed by interactions, even if the volume of the Fermi 
sea is kept fixed. Luttinger's theorem can be used to determine
the density from the volume of the Fermi sea as
$n = 2 \int \frac{d^dk}{(2\pi)^d} \, \Theta(-\txi_{\bk})$.

\subsection{Superconducting state}

For the matrix elements of the Nambu propagator we use the
standard notation
\begin{equation}
 \bG(k) = \left( \begin{array}{cc}
  G(k) & F(k) \\ F^*(k) & -G(-k)
 \end{array} \right) \; ,
\end{equation}
and the analogous expression for $\btG_0(k)$. 
The matrix elements of the self-energy are denoted by
\begin{equation}
 \btSg(k) = \left( \begin{array}{cc}
  \tSg(k) & \tS(k) \\ \tS^*(k) & -\tSg(-k)
 \end{array} \right) \; .
\end{equation}

\subsubsection{First order}

In the presence of an off-diagonal counterterm $\Delta_{\bk} \,$,
the diagonal part of $\btSg$ is still given by Eq.\ (14) to
first order, where $\tG_0(k)$ now depends on the gap function:
\begin{equation}
 \tG_0(k) = - \frac{i\om + \txi_{\bk}}
  {\om^2 + \txi_{\bk}^2 + |\Delta_{\bk}|^2} \; .
\end{equation}
The first order self-consistency relation (15) thus generalizes to
\begin{equation}
 \delta\xi_{\bk} = U \int \frac{d^dk'}{(2\pi)^d} \, 
 \frac{1}{2} \, \big( 1 - \txi_{\bk'}/E_{\bk'} \big) \; ,
\end{equation}
with $E_{\bk} = \sqrt{\txi_{\bk}^2 + |\Delta_{\bk}^2|}$.
Note that the above integral is the BCS formula for the average 
particle density per spin.

The off-diagonal matrix element of $\btSg$ is obtained from
diagrams (1a) and (1b) in Fig.\ 1 as
\begin{equation}
 \tS^{(1)}(k) = - U \int_{k'} \tF_0(k') \, + \, \Delta_{\bk} 
\end{equation}
to first order in $U$, with
\begin{equation}
 \tF_0(k) = \frac{\Delta_{\bk}}
  {\om^2 + \txi_{\bk}^2 + |\Delta_{\bk}|^2} \; .
\end{equation}
The off-diagonal part of the self-consistency condition (11) 
follows as
\begin{equation}
 \Delta_{\bk} = - U \int \frac{d^dk'}{(2\pi)^d} \, 
 \frac{\Delta_{\bk'}}{2E_{\bk'}} \; .
\end{equation}
Extended as a condition for all $\bk$ 
(and not just on $\tilde{\cal F}$)
this is nothing but the gap equation for the Hubbard model as
obtained by standard BCS theory.
The self-consistency relation requires that $\Delta_{\bk}$ be
constant on the Fermi surface, such that one naturally chooses
a constant $\Delta_{\bk} = \Delta$ as an ansatz for all $\bk$. 
A non-trivial solution $\Delta \neq 0$ of this gap equation can 
obviously be obtained only for the attractive Hubbard model 
($U < 0$). 

\subsubsection{Second order}

The diagrams (2b) and (2c) cancel each other for the same
reason as in the normal state. The contribution from diagram
(2a) to the diagonal part of the self-energy is still given
by formula (16), with $\tG_0$ from Eq.\ (20) and 
\begin{equation}
 \tPi_0(q) = - \int_{k'} \left[ 
 \tG_0(k') \, \tG_0(k'+q) + \tF_0(k') \, \tF^*_0(k'+q)
 \right] \; .
\end{equation}
The second order contribution to the off-diagonal matrix element
of $\btSg$ is
\begin{equation}
 \tS^{(2)}(k) = U^2 \int_q \tPi_0(q) \, \tF_0(k-q) \; .
\end{equation}
The self-consistency relations read:
\begin{eqnarray}
 \delta\xi_{\bk} &=& U \int \frac{d^dk'}{(2\pi)^d} \, 
 \frac{1}{2} \, \big( 1 - \txi_{\bk'}/E_{\bk'} \big) 
 \, + \, \tSg^{(2)}(0,\bk) \; , \\[2mm]
 \Delta_{\bk} &=& - U \int \frac{d^dk'}{(2\pi)^d} \, 
 \frac{\Delta_{\bk'}}{2E_{\bk'}} 
 \, - \, \tS^{(2)}(0,\bk) \; .
\end{eqnarray}
In Appendix A we present more explicit expressions for 
$\tSg^{(2)}(0,\bk)$ and $\tS^{(2)}(0,\bk)$, obtained by
carrying out the frequency integrals.

\subsection{Numerical solution}

The self-consistency conditions are non-linear equations for
the counterterms $\delta\xi_{\bk}$ and, in the superconducting
state, $\Delta_{\bk}$. The Fermi surface of the interacting
system, $\tilde{\cal F}$, on which the self-consistency conditions 
must be satisfied, is not known a priori. The equations involve 
one momentum integral at first order, and two momentum integrals
at second order. 
Such a non-linear system can only be solved iteratively.
In this subsection we describe some details of our algorithm.

Since the counterterms are determined by the self-consistency 
conditions only on the Fermi surface, their momentum dependence
away from $\tilde{\cal F}$ can be parametrized in many ways.
We have chosen $\delta\xi_{\bk}$ and $\Delta_{\bk}$ as constant
along the straight lines connecting the square shaped line 
defined by the condition $|k_x| + |k_y| = \pi$ with the points 
$(0,0)$ and $(\pi,\pi)$ of the Brillouin zone, respectively 
(see Fig.\ 2). 
For a numerical solution the remaining tangential momentum 
dependence is discretized by up to 256 points.

The iteration procedure starts with a tentative choice of 
counterterms. To be able to reach a symmetry broken solution 
one usually has to offer at least a small symmetry breaking 
counterterm in the beginning.\cite{fn1}
In each iteration step new counterterms are determined via
Eq.\ (17) in the normal state, and by Eqs.\ (27) and (28) for
the superconducting state. 
The right hand side of these equations is evaluated using the 
counterterms obtained in the previous step,
and $\bk$ is chosen on the Fermi surface defined by the 
previous $\delta\xi_{\bk}$.
The momentum integrals are carried out using a Monte-Carlo 
routine. 
The iteration is continued until convergence is achieved, that
is until the counterterms remain invariant within numerical
accuracy from step to step.
In all cases studied different choices of initial counterterms
lead to the same unique solution. 
The symmetry breaking terms are much larger than the stochastic
noise from the Monte-Carlo routine in all results shown.

The density is kept fixed by adjusting the chemical potential 
during the iteration procedure. 
To avoid a higher numerical effort we have computed the density 
from the Fermi surface volume in the normal state (justified by 
Luttinger's theorem), and from the BCS formula for the density 
in superconducting solutions. 
The latter reduces to the Fermi surface volume in the normal state 
limit, such that the potential error of this approximation is very 
small as long as the gap is small.

\section{Results}

We now discuss the most interesting results obtained within
the renormalized perturbation theory described above, focussing
mainly on the repulsive Hubbard model $(U>0$), for which we have 
found superconducting solutions with d-wave symmetry, as well as
symmetry-breaking Fermi surface deformations.

\subsection{Repulsive Hubbard model}

The following results for the repulsive Hubbard model have 
been computed for the parameters $t' = -0.15t$ and $U = 3t$.
The interaction is thus in the weak to intermediate coupling
regime. For too small $U$-values it becomes very hard to resolve 
the small superconducting gap in the numerical solution.

We have solved the self-consistency equations for various 
densities ranging from $n = 0.88$ to $n = 0.90$, for which the
Fermi surfaces are quite close to the saddle points of the 
bare dispersion relation $\eps_{\bk}$, located at $(\pi,0)$ 
and $(0,\pi)$. In all cases the normal state is unstable 
towards superconductivity. 
The gap function in the superconducting state obtained from 
the self-consistency equations has $d_{x^2-y^2}$-wave shape, 
with slight deviations from perfect d-wave symmetry in cases
where the Fermi surface breaks the symmetry of the square 
lattice.
This is in agreement with widespread expectations for the
Hubbard model,\cite{Sca} and in particular with recent 
renormalization group arguments and calculations.\cite{RG}
In Fig.\ 3 we show the gap functions obtained at the densities 
$n=0.88$ and $n=0.9$, respectively. 
We note that the size of the gap is roughly one order of magnitude
smaller than the critical cutoff scale $\Lam_c$ at which Cooper 
pair susceptibilities diverge in 1-loop renormalization group 
calculations for comparable model parameters.\cite{RG}
There are various possible reasons for this quantitative 
discrepancy.
First, and probably most importantly, the enhancement of effective 
interactions due to fluctuations, especially antiferromagnetic 
spin fluctuations, is captured much better by a renormalization 
group calculation.
Second, the approximate Fermi surface projection of vertices 
driving the renormalization group flow can lead to an 
overestimation of effective interactions and hence of critical 
energy scales.
Furthermore, a renormalization group calculation within the 
symmetry broken phase could yield a gap that is somewhat smaller 
than $\Lam_c$.

While superconductivity is the only possible instability of the
normal Fermi liquid state in the weak coupling limit (except for 
the case of perfect nesting at half-filling), at higher $U$ one 
should also consider the possibility of other, in particular 
magnetic, instabilities. 
This could be done within renormalized perturbation theory by
allowing for counterterms introducing magnetic or charge order.

The Fermi surface is always deformed by interactions. 
The shifts generated by the momentum dependence of the counterterm 
$\delta\xi_{\bk}$ are not very large. 
They are more pronounced near the saddle points of $\eps_{\bk}$, 
where small energy shifts lead to relatively large shifts in 
k-space. 
However, the results presented in Fig.\ 4 show that the Fermi 
surface of the interacting system can nevertheless differ 
strikingly from the bare one.
For the densities $n=0.88 - 0.889$ the Fermi surface of the 
interacting system obviously breaks the point group symmetry of
the square lattice. 
For $n=0.88$ and $n=0.888$ even the topology of the Fermi surface
is changed by interactions. The deformed surface has open 
topology in these cases, instead of being closed around the points 
$(0,0)$ or $(\pi,\pi)$ in the Brillouin zone. 
Note that the symmetry-broken Fermi surfaces shown here correspond
to stable solutions of the self-consistency equations for the 
counterterms, while symmetric solutions are unstable.

More details about the Fermi surface shifts can be extracted 
from a plot of the second order counterterms, shown in Fig.\ 5.
The actual shifts are determined by these terms plus a constant
due to the first order counterterm and a shift of the chemical 
potential.
At fixed density the interaction shifts the Fermi surface 
outwards at points where $\tSg(0,\tilde\bk_F)$ has an absolute 
minimum, and inwards at points corresponding to absolute maxima.
Interactions thus reduce the curvature of the Fermi surface near
the diagonals in the Brillouin zone. 
Fig.\ 5 reveals that the Fermi surface deformation is slightly 
asymmetric also for $n=0.9$, but the symmetry breaking is too
small to be seen in Fig.\ 4. 

If the Fermi surface breaks the square lattice symmetry, the gap 
function $\Delta_{\bk}$ cannot have pure d-wave symmetry any more. 
See, for example, the gap function at density $n=0.88$ in Fig.\ 3.  
The deviation from perfect d-wave form is however quite small,
since the symmetry breaking Fermi surface deformation is small.

Interaction-induced Fermi surface deformations which break the
symmetry of the square lattice have already been discussed 
earlier in the literature. 
Yamase and Kohno \cite{YK} have obtained symmetry-broken Fermi 
surfaces within a slave boson mean-field theory for the t-J model. 
The effective interactions obtained from 1-loop renormalization
group flows for the Hubbard model also favor symmetry-breaking
Pomeranchuk instabilities of the Fermi surface, if the latter
is close to the van Hove points.\cite{HM}
A systematic stability analysis of the Hubbard model using
Wegner's Hamiltonian flow equation method confirmed that
symmetry breaking Fermi surface deformations are among the 
strongest instabilities.\cite{GKW}
It remained an open question, however, whether such Fermi surface 
instabilities would be cut off by the superconducting gap. 
We have observed within our renormalized perturbation theory 
that symmetry breaking Fermi surface deformations occur indeed
more easily, if the system is forced to stay in a normal state,
by setting $\Delta_{\bk} = 0$.
Whether a symmetry broken Fermi surface and superconductivity 
coexist can be seen only by performing a calculation within the 
symmetry-broken state. 
This has not yet been done using the renormalization group or
flow equation methods. 

From a pure symmetry-group point of view the symmetry breaking 
generated by the Pomeranchuk instability is equivalent to that 
in ''nematic'' electron liquids, first discussed by Kivelson et 
al.\cite{KFE}. These authors considered doped Mott insulators, 
that is {\em strongly} interacting systems. 
A general theory of orientational symmetry-breaking in fully 
isotropic (not lattice) two- and three-dimensional Fermi liquids 
has been reported by Oganesyan et al.\cite{OKF}
Superconducting nematic states, in which discrete orientational 
symmetry breaking develops in addition to d-wave superconductivity, 
have been considered recently by Vojta et al.\cite{VZS} 
Motivated by experimental properties of single-particle excitations 
in cuprate superconductors they performed a general classification 
and field-theoretic analysis of various phases with an additional 
order parameter on top of $d_{x^2-y^2}$-pairing.

\subsection{Attractive Hubbard model}

For the attractive Hubbard model ($U<0$) the renormalized 
perturbation expansion yields s-wave superconductivity already 
at first order, which is equivalent to BCS mean-field 
theory.\cite{MRR}
At this level the gap function is constant in k-space.
Extending the calculation to second order, a weak momentum
dependence of $\Delta_{\bk}$ is generated, as seen in Fig.\ 6
for the parameters $U = -2t$, $t' = -0.15t$ and $n=0.9$.
More importantly, the overall size of the gap is strongly reduced 
by fluctuations included in the second order terms. 
The average gap in Fig.\ 6 is only one third of the corresponding 
mean-field gap.
It has been pointed out previously that fluctuations not contained
in mean-field theory reduce the size of magnetic and other order 
parameters even in the weak coupling limit.\cite{GY,Don}

\section{Conclusion}

In summary, we have formulated a renormalized perturbation theory
for interacting Fermi systems, which treats Fermi surface 
deformations and superconductivity via additive counterterms. 
This method is very convenient for studying the role of 
fluctuations for spontaneous symmetry breaking in a controlled
weak-coupling expansion. 
A concrete application of the expansion carried out to second 
order yields several non-trivial results for the two-dimensional 
Hubbard model. In particular, for the repulsive model we have 
obtained the gap function of the expected d-wave superconducting 
state and, for Fermi levels close to the van Hove energy, an 
interacting Fermi surface with broken lattice symmetry, and in
some cases even open topology.
The symmetry-breaking pattern of the states with symmetry-broken 
Fermi surfaces is equivalent to that of ''nematic'' electron
liquids discussed already earlier from a different point of
view.\cite{KFE,VZS}

The present work can be extended in several interesting
directions.
After fixing the counterterms one can compute the full 
momentum and energy dependence of the self-energy, and hence the 
spectral function for single-particle excitations. At second order 
the combined effects of symmetry breaking and quasi-particle decay
are captured.
Allowing for other symmetry-breaking counterterms, for example
spin density waves, one can study the competition of magnetic,
charge, and superconducting instabilities, as well as their 
possible coexistence. 
Finally, the formalism can be extended to finite temperature.
In that case the singularities of the bare propagator are cut off
by the smallest Matsubara frequency, but Fermi surface shifts and
symmetry breaking can still be conveniently taken into account
by counterterms.

\vskip 1cm

\noindent
{\bf Acknowledgements:} \\
We are grateful to A.\ Georges, M.\ Keller, D.\ Rohe and 
M.\ Salmhofer for valuable discussions, and to D.\ Rohe also for
a critical reading of the manuscript.

\begin{appendix}

\section{Frequency integrals} 

The Matsubara frequency integrals in the second order self-energy
contributions can be carried out analytically by using the
residue theorem. 
We only present the results for the superconducting case; 
the normal state results can be recovered by setting 
$\Delta_{\bk} = 0$ in the following expressions.

The frequency integrals relevant for the evaluation of $\tPi_0$
defined by Eq.\ (25) are
\begin{eqnarray}
 \int \frac{dk_0}{2\pi} \, \tG_0(k) \, \tG_0(k+q) & = &
 \frac{E_{\bk} + E_{\bk+\bq}}{2 E_{\bk} E_{\bk+\bq}} \,
 \frac{\txi_{\bk} \txi_{\bk+\bq} - E_{\bk} E_{\bk+\bq}}
  {q_0^2 + [E_{\bk} + E_{\bk+\bq}]^2} \nonumber \\
 & + &
 \frac{iq_0}{2 E_{\bk} E_{\bk+\bq}} \,
 \frac{\txi_{\bk} E_{\bk+\bq} - E_{\bk} \txi_{\bk+\bq}}
  {q_0^2 + [E_{\bk} + E_{\bk+\bq}]^2}
\end{eqnarray}
and
\begin{equation}
 \int \frac{dk_0}{2\pi} \, \tF_0(k) \, \tF^*_0(k+q) =
 \frac{E_{\bk} + E_{\bk+\bq}}{2 E_{\bk} E_{\bk+\bq}} \,
 \frac{\Delta_{\bk} \Delta^*_{\bk+\bq}}
  {q_0^2 + [E_{\bk} + E_{\bk+\bq}]^2} \; .
\end{equation}
The imaginary part of $\tPi_0$ does not contribute to
$\btSg(0,\bk)$. Carrying out the $q_0$-integral in Eqs.\ (16)
and (26) yields:
\begin{eqnarray}
 \tSg^{(2)}(0,\bk) & = & - U^2 \int_{\bq} \int_{\bk'}
 \txi_{\bk-\bq} \, C(\bk,\bk',\bq) \\[2mm]
 \tS^{(2)}(0,\bk)  & = & U^2 \int_{\bq} \int_{\bk'}
 \Delta_{\bk-\bq} \, C(\bk,\bk',\bq) \; ,
\end{eqnarray}
where
\begin{equation}
 C(\bk,\bk',\bq) = 
 \frac{E_{\bk'} E_{\bk'+\bq} - \txi_{\bk'} \txi_{\bk'+\bq} - 
  \Delta_{\bk'} \Delta^*_{\bk'+\bq}}
  {4 E_{\bk-\bq} E_{\bk'} E_{\bk'+\bq} \,
  \big[ E_{\bk-\bq} + E_{\bk'} + E_{\bk'+\bq} \big]} \; .
\end{equation}

\end{appendix}


\vfill\eject


\centerline{\Large FIGURES}
\vskip 1cm

\begin{figure}
\center
\epsfig{file=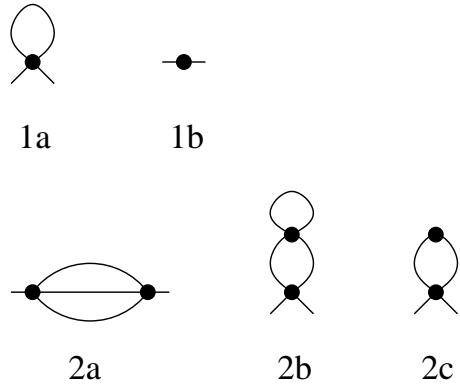,width=7cm}
\caption{The Feynman diagrams contributing to the renormalized
 self-energy $\tSg$ at first and second order perturbation theory;
 the two-particle vertices represent the antisymmetrized 
 interaction, one-particle vertices the counterterm,
 and lines the renormalized bare propagator $\tG_0$.}
\label{fig:fig1}
\end{figure}

\vskip 1cm

\begin{figure}
\center
\epsfig{file=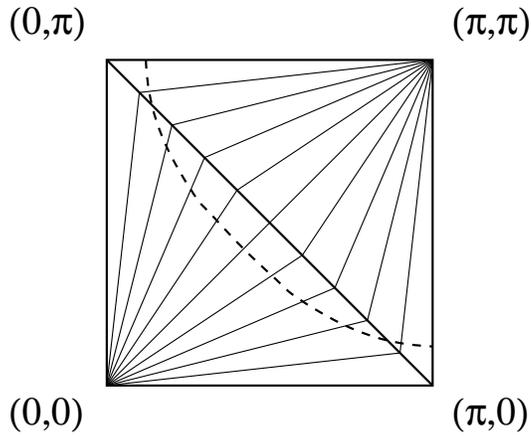,width=7cm}
\vskip 1cm
\caption{Parametrization of counterterms in the first quadrant 
 of the Brillouin zone; the counterterms are constant along the 
 straight lines connecting the line from $(\pi,0)$ to $(0,\pi)$ 
 with the points $(0,0)$ and $(\pi,\pi)$, respectively; 
 the dashed line illustrates a typical Fermi surface.}
\label{fig:fig2}
\end{figure}

\vfill\eject

\begin{figure}
\center
\epsfig{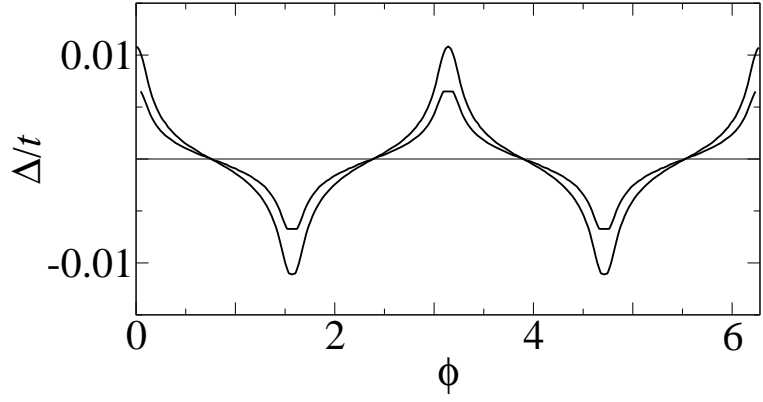}
\vskip 1cm
\caption{Gap function for $n=0.88$ (larger amplitude) and $n=0.9$
 (smaller amplitude) as a function of the angle with respect to 
 the $k_x$-axis.}
\label{fig:fig3}
\end{figure}

\vskip 1cm

\begin{figure}
\center
\epsfig{file=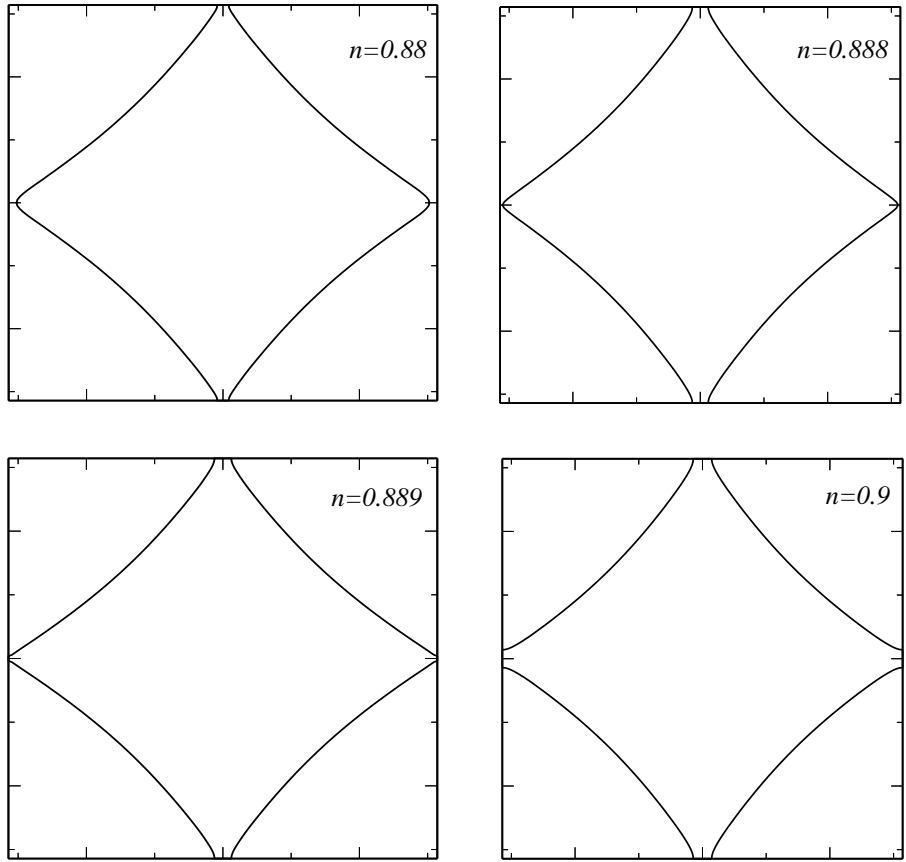,width=12cm}
\vskip 1cm
\caption{Fermi surfaces of the interacting system for 
 different densities $n$.}
\label{fig:fig4}
\end{figure}

\vfill\eject

\begin{figure}
\center
\epsfig{file=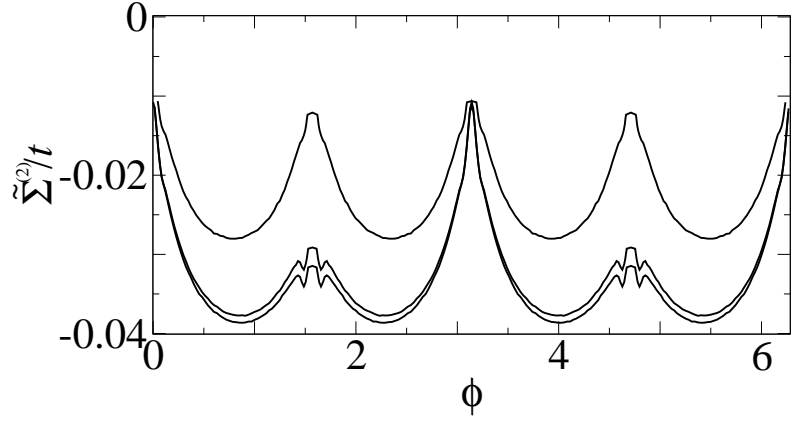,width=12cm}
\vskip 1cm
\caption{Second-order counterterms $\tSg^{(2)}(0,\tilde\bk_F)$ 
 as a function of the angle with respect to the $k_x$-axis, 
 for the densities $n = 0.888, 0.889, 0.9$ (from bottom to top).}
\label{fig:fig5}
\end{figure}

\vskip 2cm

\begin{figure}
\center
\epsfig{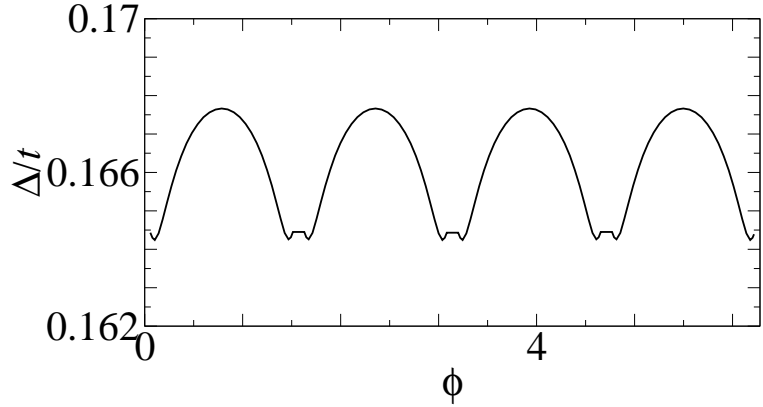}
\vskip 1cm
\caption{Gap function for $n = 0.9$ as a function of the angle 
 with respect to the $k_x$-axis for the attractive ($U=-2t$) 
 Hubbard model.}
\label{fig:fig6}
\end{figure}


\begin{thebibliography}{99}
\bibitem{fn1} Also small numerical errors can be sufficient
 to start the symmetry breaking in the iteration procedure.
\end{thebibliography}
\end{document}